\documentclass[apjl]{emulateapj}

\slugcomment{ 1/July/2014 }

\shorttitle{Insights on the stellar mass-metallicity relation from the CALIFA survey}
\shortauthors{Gonz\'alez Delgado \& the CALIFA collaboration}

\usepackage[usenames]{color}
\newcommand\starlight{{\sc starlight}}          	% STARLIGHT 
\begin{document}


\title{Insights on the stellar mass-metallicity relation from the CALIFA survey}

\author{R. M. Gonz\'alez Delgado\altaffilmark{1}}
\author{R. Cid Fernandes\altaffilmark{2}}
\author{R. Garc\'\i a-Benito\altaffilmark{1}}
\author{E. P\'erez\altaffilmark{1} }
\author{A. L. de Amorim\altaffilmark{2} }
\author{C. Cortijo-Ferrero\altaffilmark{1}}
\author{E. A. D. Lacerda\altaffilmark{2} }
\author{R. L\'opez Fern\'andez\altaffilmark{1} }
\author{S. F. S\'anchez\altaffilmark{1,3,4}}
\author{N. Vale Asari\altaffilmark{2} }
\author{J. Alves\altaffilmark{5} }
\author{J. Bland-Hawthorn\altaffilmark{6} }
\author{L. Galbany\altaffilmark{7} }
\author{A. Gallazzi\altaffilmark{8, 19} }
\author{B. Husemann\altaffilmark{9, 10} }
\author{S. Bekeraite\altaffilmark{10} }
\author{B. Jungwiert\altaffilmark{11} }
\author{A. R. L\'opez-S\'anchez\altaffilmark{12} }
\author{A. de Lorenzo-C\'aceres\altaffilmark{13} }
\author{R. A. Marino\altaffilmark{14}}
\author{D. Mast\altaffilmark{15}}
\author{M. Moll\'a\altaffilmark{16} }
\author{A. del Olmo\altaffilmark{1} }
\author{P. S\'anchez-Bl\'azquez\altaffilmark{17} }
\author{G. van de Ven\altaffilmark{18} }
\author{J. M. V\'\i lchez\altaffilmark{1} }
\author{C. J. Walcher\altaffilmark{10} }
\author{L. Wisotzki\altaffilmark{10} }
\author{B. Ziegler\altaffilmark{5} }
\author{CALIFA collaboration \altaffilmark{20}}

\affil{\altaffilmark{1}
Instituto de Astrof\'\i sica de Andaluc\'\i a (CSIC), Glorieta de la Astronom\'\i a s/n, E-18008 Granada, Spain.}
\affil{\altaffilmark{2}
Departamento de F\'\i sica, Universidade Federal de Santa Catarina, P.O. Box 476, 88040-900, Florian\'opolis, SC, Brazil}
\affil{\altaffilmark{3}
Centro Astron\'omico Hispano Alem\'an, Calar Alto, (CSIC-MPG), Jes\'us Durb\'an Rem\'on 2-2, E-04004 Almer\'\i a, Spain}
\affil{\altaffilmark{4}
 Instituto de Astronom\'\i a,Universidad Nacional Auton\'oma
de Mexico, A.P. 70-264, 04510, M\'exico,D.F.}
\affil{\altaffilmark{5}
University of Vienna, T\"urkenschanzstrasse 17, 1180, Vienna, Austria}
\affil{\altaffilmark{6} 
Sydney Institute for Astronomy, The University of Sydney, NSW 2006, Australia}
\affil{\altaffilmark{7}  
Millennium Institute of Astrophysics and Departamento de Astronom\'\i a, Universidad de Chile, Casilla 36-D, Santiago, Chile}
\affil{\altaffilmark{8}
INAF $-$ Osservatorio Astrofisico di Arcetri, Largo Enrico Fermi 5,
50125 Firenze, Italy }
\affil{\altaffilmark{9} 
European Southern Observatory, Karl-Schwarzschild-Str. 2, 85748 Garching b. M\"unchen, Germany}
\affil{\altaffilmark{10} 
Leibniz-Institut f\"ur Astrophysik Potsdam, An der Sternwarte 16, D-14482 Potsdam, Germany}
\affil{\altaffilmark{11} 
Astronomical Institute of the Academy of Sciences of the Czech Republic, v.v.i., Bocni II 1401, 14131 Prague, Czech Republic}
\affil{\altaffilmark{12}
Australian Astronomical Observatory, PO BOX 296, Epping, 1710
NSW, Australia}
\affil{\altaffilmark{13}
School of Physics and Astronomy, University of St. Andrews, North Haugh, St. Andrews, KY169SS, UK}
\affil{\altaffilmark{14}
CEI Campus Moncloa, UCM-UPM, Departamento de Astrof\'{i}sica y CC. de la Atm\'{o}sfera, Facultad de CC.\ F\'{i}sicas, Universidad Complutense de Madrid, Avda.\,Complutense s/n, 28040 Madrid, Spain}
\affil{\altaffilmark{15} 
Instituto de Cosmologia, Relatividade e Astrof\'{i}sica Ð ICRA, Centro Brasileiro de Pesquisas F\'{i}sicas, Rua Dr.Xavier Sigaud 150, CEP 22290-180, Rio de Janeiro, RJ, Brazil }
\affil{\altaffilmark{16} 
Departamento de Investigaci\'on B\'asica, CIEMAT, Avda.\ Complutense 40, E-28040 Madrid, Spain}
\affil{\altaffilmark{17} 
Depto. de F\'{\i}sica Te\'orica, Universidad Aut\'onoma de Madrid, 28049 Madrid, Spain}
\affil{\altaffilmark{18} 
Max-Planck-Institut f\"ur Astronomie, K\"onigstuhl 17, D-69117 Heidelberg, Germany}
\affil{\altaffilmark{19}
Dark Cosmology Center, University of Copenhagen, Niels Bohr Institute, Juliane Maries Vej 30, 2100 Copenhagen, Denmark}
\affil{\altaffilmark{20}
CALIFA International Collaboration \url{http://califa.caha.es}}

\begin{abstract}

We use spatially and temporally resolved maps of stellar population properties of 300 galaxies from the CALIFA integral field survey to investigate how the stellar metallicity ($Z_\star$) relates to the total stellar mass ($M_\star$) and the local mass surface density ($\mu_\star$) in both spheroidal and disk dominated galaxies.
The galaxies are shown to follow a clear stellar mass-metallicity relation (MZR) over the whole $10^9$ to $10^{12} M_\odot$ range. This relation is steeper than the one derived from nebular abundances, which is similar to the flatter stellar MZR derived when we consider only young stars. 
We also find a strong relation between the local values of  $\mu_\star$ and $Z_\star$ (the $\mu$ZR), betraying the influence of local factors in determining $Z_\star$.
This shows that both local ($\mu_\star$-driven) and global ($M_\star$-driven) processes are important in determining the metallicity in galaxies.
We find that the overall balance between local and global effects varies with the location within a galaxy. In disks, $\mu_\star$ regulates $Z_\star$, producing a strong $\mu$ZR whose amplitude is modulated by $M_\star$. 
In spheroids it is $M_\star$ who dominates the physics of star formation and chemical enrichment, with $\mu_\star$ playing a minor, secondary role. These findings agree with our previous analysis of the star formation histories of CALIFA galaxies, which showed that mean stellar ages are mainly governed by surface density in galaxy disks and by total mass in spheroids.

\end{abstract}


\section{Introduction}

\label{sec:Intro}

The quest to understand the physical association between mass and metallicity in galaxies has a long history in astrophysics.
In its currently most common form, mass is represented by the stellar mass ($M_\star$), while the nebular oxygen abundance is taken as the metallicity tracer. Perhaps the best known contemporary example of this mass-metallicity relation (MZR) is that derived for star-forming galaxies in the SDSS by Tremonti et al.\ (2004).

An independent way to explore the metal content of galaxies is through their stars, whose combined spectra encode a fossil record of galaxy evolution. This approach has been 
widely explored in relation to elliptical galaxies, which also follow a MZR (the Mg-$\sigma_\star$ relation; Faber \& Jackson 1976). The historical limitation to early type systems comes about because of their predominantly old and relatively simple stellar populations (often approximated as a single burst), in contrast with the more complex star formation histories (SFH) of spirals, which complicates the translation of their spectroscopic features into tracers of the characteristic stellar metallicity ($Z_\star$). Recent work has started to lift this restriction. In particular, spectral synthesis techniques have shown that useful estimates of $Z_\star$ can be derived for galaxies of all types (Cid Fernandes et al.\ 2005, 2007; Gallazzi et al.\ 2005; Panter et al.\ 2007), especially when used  comparatively and for large samples. 
These studies obtain a purely stellar version of the MZR, as well as a broad correlation between O/H and $Z_\star$, and a $Z_\star$ evolutionary pattern characteristic of chemical enrichment, all derived on the basis of an archeological spectral analysis of local SDSS galaxies.

A common problem to both approaches is that they are most often based on integrated data, where one gathers all or part of the light of a galaxy in a single spectrum. The exact meaning of the global O/H, $Z_\star$, and other properties derived from such data is rendered somewhat ambiguous (and aperture dependent; e.g., Iglesias-P\'aramo et al.\ 2014; Mast et al.\ 2014) by the existence of spatial gradients in both nebular and stellar population properties within galaxies. More critically, the lack of spatial resolution limits our ability to map the influence of local factors on the local metallicity which, if present, propagate to the global MZR.

Integral field spectroscopy surveys like CALIFA (S\'anchez et al.\ 2012) best tackle these issues. 
Rosales-Ortega et al.\ (2012) and S\'anchez et al.\ (2013) applied nebular diagnostics to thousands of {H\,{\sc ii}} regions in these datacubes, finding that O/H is strongly correlated with the local stellar mass {\em surface density} ($\mu_\star$), and that this relation shapes the global MZR. The notion that $M_\star$ plays a less fundamental role than $\mu_\star$ in the evolution of galactic disks (home of the {H\,{\sc ii}} regions whose O/H represent the metallicity in the nebular MZR) has been raised previously. Bell \& de Jong (2000), for instance, analysed optical and near-IR imaging of spirals, finding that the surface density of a galaxy drives its SFH, $M_\star$ being a less important parameter. In the context of integral field studies, Gonz\'alez Delgado et al.\ (2014, hereafter GD14) carried out a spectral synthesis analysis of 107 CALIFA galaxies, also finding that mean stellar ages correlate more strongly with $\mu_\star$ in galaxy disks. Conversely, they find that in spheroids (elliptical galaxies and bulges), it is $M_\star$ which controls the SFH.

Here we explore global and local estimates of the $Z_\star$ obtained from the spatially resolved spectral synthesis of CALIFA galaxies. The focus on stars complements the work based on {H\,{\sc ii}} regions, bringing in new perspectives. First, stellar metallicities reflect the whole history of a galaxy, while O/H measures a present-day snapshot of its evolution. Second, our analysis ignores {H\,{\sc ii}} regions, and so can be applied to disks and spheroids, increasing statistics and broadening the scope of the study.

\section{Data and stellar population analysis}

\label{sec:Method}

Our sample comprises 300 galaxies ranging from ellipticals to late type spirals, representative of the full CALIFA mother sample (Walcher et al.\ 2014). 
Technical details are described in S\'anchez et al.\ (2012), Husemann et al.\ (2013), and the forthcoming Data Realease 2 article by Garc\'\i a-Benito et al.\ (in preparation).

Our method to extract stellar population properties from datacubes has been explained in P\'erez et al.\ (2013), Cid Fernandes et al.\ (2013, 2014). In short, we use
\starlight\ (Cid Fernandes et al.\ 2005) to fit each spectrum as a combination of simple stellar populations (SSPs) spanning different ages and metallicities. 
The SSP spectra used in this work consist of Granada (Gonz\'alez Delgado et al.\ 2005) and MILES (Vazdekis et al.\ 2010) models, similar to those used in Cid Fernandes et al.\ (2013) but extended to cover the full metallicity range of the MILES models. The 235 elements base covers ages from 0.001 to 14 Gyr, and $\log Z_\star/Z_\odot$ from $-2.3$ to $+0.33$. A Salpeter initial mass function is adopted.

This whole analysis transforms the datacubes to 2D maps of a series of stellar population properties. The main ones for the purposes of this Letter are the stellar mass surface density, and the mass-weighted mean stellar metallicity, defined as

\begin{equation}
\label{eq:logZmass}
\langle \log Z_\star \rangle_{M,xy} = 
\frac{ \sum_{tZ} M_{\star,tZ,xy} \times \log\ Z}{
\sum_{tZ} M_{\star,tZ,xy} }
\end{equation}

\noindent  where $M_{\star,tZ,xy}$ denotes the mass in stars of age $t$ and metallicity $Z$ in spaxel $xy$. 

\section{Results}

\subsection{The global mass metallicity relation}

%***FIG***FIG***FIG***FIG***FIG***FIG***FIG***FIG***FIG***FIG***
%\begin{figure*}[!ht]
\begin{figure*}
\includegraphics[width=0.9\textwidth]{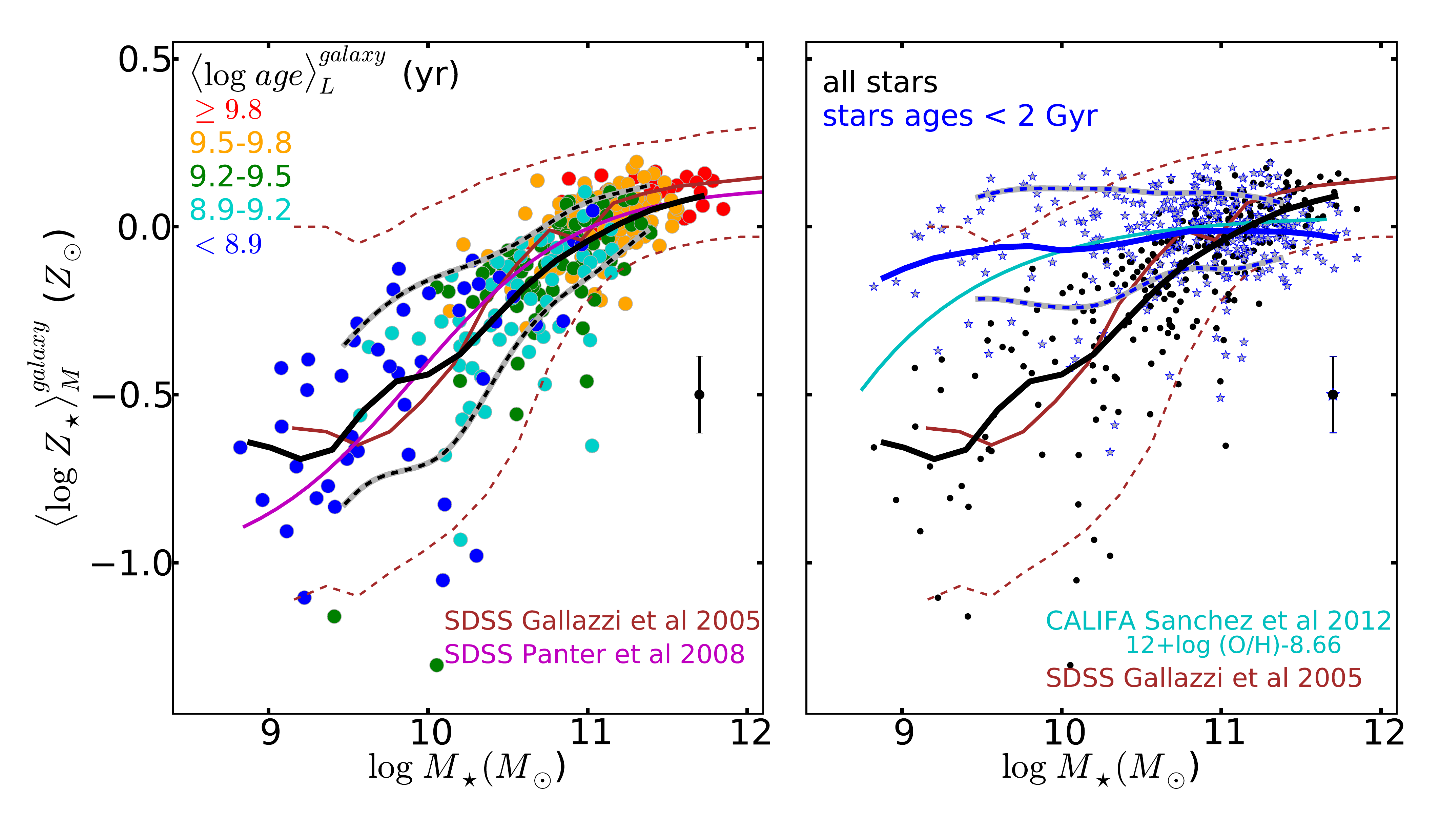}
\caption{{\em Left:} The global {\em stellar} MZR for 300 CALIFA galaxies is shown as circles, color coded by the mean stellar age. A mass-binned smooth mean relation is shown as a black solid line (the dashed black indicates the 16 and 84 percentiles, and the error bar represents the typical dispersion). The MZRs obtained for SDSS galaxies by Gallazzi et al.\ (2005) and Panter et al.\ (2008) are plotted as brown and magenta lines, respectively, with dashed brown indicating the 16 and 84 percentiles of Gallazzi et al. 
{\em Right:} Comparison of our global stellar MZR (black circles and line) with the one obtained considering only stars younger than 2 Gyr in the computation of the stellar metallicity (blue stars and line); blue dashed lines indicate the 16 and 84 percentiles;  error bar represents the typical dispersion. The CALIFA-based {\em nebular} MZR (S\'anchez et al.\ 2013) is shown as a cyan line. }
\label{fig:MZR1}
\end{figure*}
%***FIG***FIG***FIG***FIG***FIG***FIG***FIG***FIG***FIG***FIG***

The MZR is a relation between {\em global} galaxy properties, which requires us to compress our $xy$ maps to single numbers. The total stellar mass is obtained from $M_\star = \sum_{xy} M_{\star,xy}$, i.e., adding up the masses derived for individual spaxels (thus accounting for spatial variations in the mass-to-light ratio and extinction). For the metallicity we use a galaxy-wide averaging scheme, such that

\begin{equation}
\label{eq:logZmass_Galaxy}
\langle \log Z_\star \rangle_M^{galaxy} = 
\frac{ \sum_{xy} M_{\star,xy} \langle \log Z_\star \rangle_{M,xy} }{ \sum_{xy} M_{\star,xy} }
\end{equation}

\noindent as done for light and mass weighted mean stellar ages
by GD14, who showed that these averages match very well the value obtained from the synthesis of 
the integrated spectrum, as well as that at radial distances of $R = 1$ Half Light Radius (HLR).

Our mass vs.\ stellar metallicity relation for CALIFA galaxies is shown in Fig.\ \ref{fig:MZR1}a.  A clear correlation is seen, with metallicities rising by roughly an order of magnitude for $M_\star$ between $\sim 10^9$ and $10^{12} M_\odot$. 
The thick black curve, obtained by smoothing the mass-binned relation, offers as better visualization of the general trend. In order to get a sense of the SFHs in this plane, we color-code the points by the value of the galaxy-wide mean stellar age (cf.\ Eq.\ 1 of GD14). One sees that galaxies become progressively older as one moves up the MZR.

Fig.\ \ref{fig:MZR1} also compares our stellar MZR with those obtained for SDSS galaxies by Gallazzi et al.\ (2005, brown line) and Panter et al.\ (2008, magenta), both shifted by $+0.25$ dex in mass to match our initial mass function. Given the significant differences in data, samples, and methodology, one should not take this comparison too literally. The overall agreement, however, is remarkable.

\subsection{Stellar vs.\ nebular MZRs and chemical evolution}

The flattening of our MZR at $\sim 10^{11} M_\odot$ is reminiscent of the behaviour observed in the nebular MZR (e.g., Tremonti et al.\ 2004). A quantitative comparison of nebular and stellar metallicities is not warranted given the huge differences in the underlying physics, not to mention inherent uncertainties affecting both estimates. 
A qualitative comparison, however, is instructive. Fig.\ \ref{fig:MZR1}b overplots the S\'anchez et al.\ (2013, cyan line) relation between $M_\star$ and O/H obtained from {H\,{\sc ii}} regions in CALIFA datacubes. The two (black and cyan lines) occupy the same zone at high $M_\star$, but diverge at low $M_\star$, with the nebular MZR being visibly flatter than the stellar one. 

Besides the caveats noted above, this comparison is skewed by the fact that O/H portrays the current state of the warm gas, while $Z_\star$ reflects the whole history of a galaxy. To mitigate this mismatch in time-scales we compute $Z_\star$ considering only stars younger than 2 Gyr, and derive the corresponding galaxy-wide average.
% (Eq.\ \ref{eq:logZmass_Galaxy}). 
The resulting MZR is shown as blue stars in Fig.\ \ref{fig:MZR1}b, with a thick blue line representing its smoothed version. 

The plot reveals a clear signature of chemical evolution, with young stars being more enriched than older stars.
% (if $M_\star \leq 10^{11} M_\odot$). 
The MZR for young stars is flatter than the global one, in line with the nebular relation. 

The overall picture portrayed by Fig.\ \ref{fig:MZR1} is the same as that drawn by Cid Fernandes et al.\ (2007) in their analysis of SDSS galaxies: Most of the chemical evolution of massive galaxies has occurred many Gyr ago, while low $M_\star$ systems grow their metallicities over a longer cosmic time span. This ``chemical downsizing'' is, of course, not coincidental, since star formation and chemical evolution work in tandem.

\subsection{The local and global $\mu_\star$ vs.\ $Z_\star$ relations}

%***FIG***FIG***FIG***FIG***FIG***FIG***FIG***FIG***FIG***FIG***
\begin{figure*}
\includegraphics[width=\textwidth]{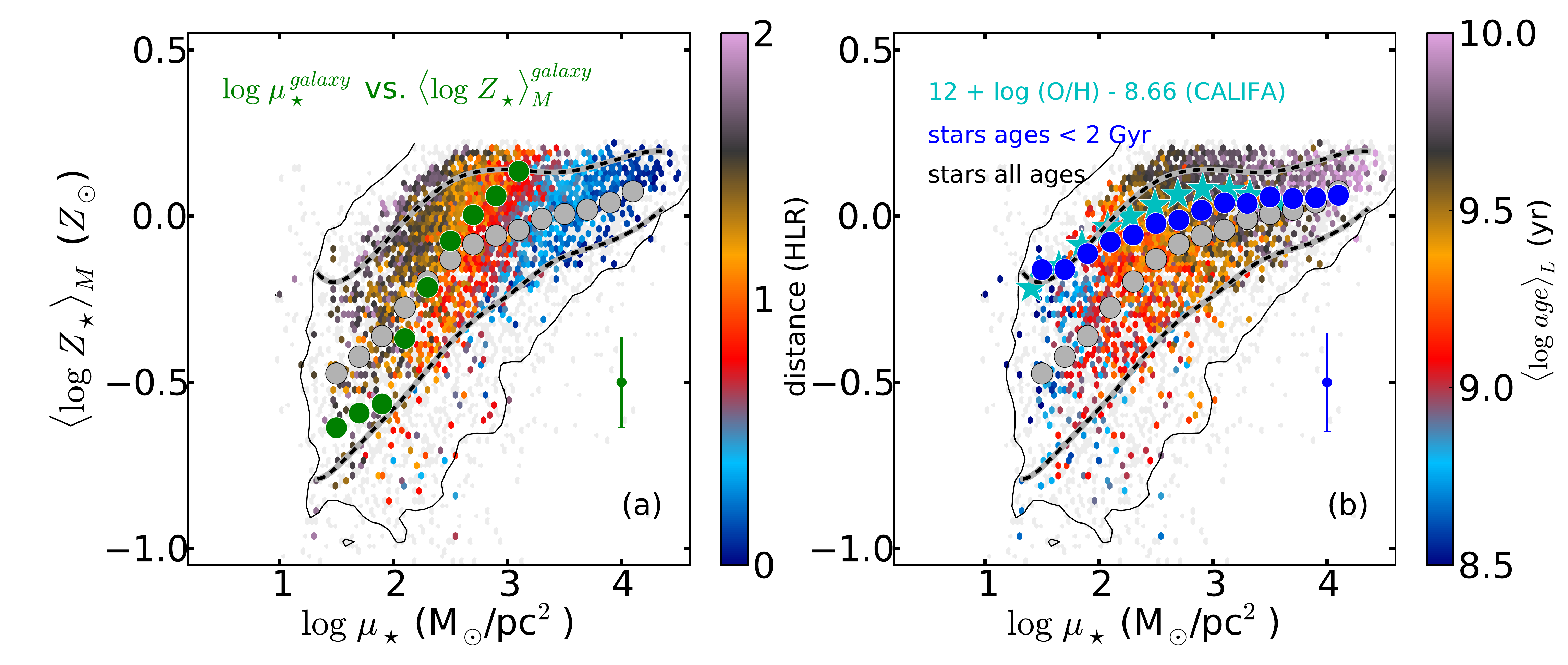}
\caption{Local stellar metallicity versus the local stellar mass surface density. Small  dots show values for 6000 radial bins of 300 galaxies. The solid line traces the contour of point number density larger than 1 per 0.075$\times$0.035 dex box.
{\em Left:} Gray circles track the $\mu_\star$-binned average $Z_\star(\mu_\star)$ relation (black dashed lines indicate the 16 and 84 percentiles).
Green circles trace the ($\mu_\star$-binned) {\em global} relation, obtained from the $\langle \log Z_\star \rangle_M^{galaxy}$ and $\mu_\star^{galaxy}$ galaxy-wide averages; the error bar represents the typical dispersion in this relation. The individual points are color-coded by the distance from the center (in HLR). 
{\em Right:} Dots as in panel (a), but colored to display the (luminosity weighted) mean stellar age. Large blue circles show the $\mu_\star$-binned global relation obtained considering only stars younger than 2 Gyr in the computation of $Z_\star$ (error bar represents the typical dispersion). Cyan stars show the CALIFA-based {\em nebular} $\mu$ZR of S\'anchez et al.\ (2013), assuming  $12+ \log (O/H)_\odot = 8.66$.
}
\label{fig:muZR1}
\end{figure*}
%***FIG***FIG***FIG***FIG***FIG***FIG***FIG***FIG***FIG***FIG***

Our data are ideally suited to investigate the relative roles of global and local properties in controlling the metallicity. To explore the information encoded in our spatially resolved maps of stellar population properties we first compress them to 1D radial profiles computed as explained in GD14. Besides allowing for clearer visualization of the results, this compression {\em (i)} reduces statistical uncertainties (see Cid Fernandes et al.\ 2013), and {\em (ii)} balances the role of each galaxy in our statistics, which would otherwise be biased towards galaxies filling up more spaxels in the integral field unit.

Radial profiles  of $\mu_\star$ and $\langle \log Z_\star \rangle_M$ were computed in elliptical annuli in steps of 0.1 HLR. The profiles typically cover out to $R = 2$ HLR reliably, so we restrict our analysis to this limit.

Fig.\ \ref{fig:muZR1}a shows the resulting $\mu_\star$-$Z_\star$ relation ($\mu$ZR). Note that this plot is ultimately a collection of 300 metallicity profiles, where the radial coordinate is replaced by $\mu_\star(R)$, such that $R$ increases to the left. The $\sim 6000$ dots (20 per galaxy) are color coded according to their distance from the nucleus, in units of the corresponding HLR. Points in the upper right correspond to the inner regions of massive, early type spheroids, while on the bottom left the outer zones of late type spirals dominate.

Clearly, despite the scatter, local stellar mass surface density and metallicity are strongly correlated, and in a radially dependent fashion. The general trend is depicted as gray circles, which averages over metallicities in 0.2 dex wide bins in $\mu_\star$. 

The average $\mu$ZR  traced by the gray circles in Fig.\ \ref{fig:muZR1}a should not be confused with the global one, since $R$-bins do not contribute equally to the galaxy-wide averages of neither $\mu_\star$ nor $\langle \log Z_\star \rangle_M$. The  ($\mu_\star$-binned) global version of the $\mu$ZR is shown by the large green circles. Note that they follow closely the underlying orange/red points ($R \sim 1$ HLR in our color scheme), as expected from the fact that galaxy-wide averages essentially reflect properties at 1 HLR (GD14).

The global $\mu$ZR is visibly steeper than the local one, particularly in the high density regime of the inner regions of galaxies (blueish points in Fig.\ \ref{fig:muZR1}a), where the local $\mu$ZR (gray circles) flattens. This flattening is not picked up by the global relation, as there are no galaxies globally as dense as these central regions of spheroids.

Overall, we conclude that the $\mu$ZR shows the local stellar surface density to be an important driver of the stellar metallicity, except for their innermost densest regions, where some other property must take over the dominant role. Even in the outer regions, however, the vertical scatter in the $\mu$ZR is such that $\mu_\star$ cannot be the sole parameter regulating $Z_\star$.

\subsection{Stellar vs.\ nebular $\mu$ZRs and chemical evolution}

Before furthering our investigation of global and local effects, we repeat in Fig.\ \ref{fig:muZR1}b the exercise carried out in 
Fig.\ \ref{fig:MZR1}b for the global MZR, but now for the $\mu$ZR. The large gray circles are as in panel (a), but the blue ones track the relation obtained considering only young stars. Again one obtains a flatter relation, indicative of chemical evolution. Also, as in Fig.\ \ref{fig:MZR1}b, the shape of the $\mu$ZR for young stellar  populations approaches that obtained from {H\,{\sc ii}} regions (cyan-colored stars, S\'anchez et al.\ 2013).

Fig.\ \ref{fig:muZR1}b also summarizes how local SFHs vary across the $\mu$Z plane. This is revealed by the colors of the dots, which now indicate the radially-averaged mean stellar age. As for the global MZR, one finds younger systems 
(i.e., those with more significant recent star-formation)
towards the low $\mu_\star$, metal poor corner of the $\mu$ZR. Large ages, on the other hand, occur at the densest, more metal rich regions typical of small $R$.
These inner zones essentially completed their star formation and chemical evolution long ago. Notice that this is also the regime where the local $\mu$ZR (gray circles) flattens.

\section{Local $+$ global effects and the distinct roles of 
$\mu_\star$ and $M_\star$ in disks and spheroids}

%***FIG***FIG***FIG***FIG***FIG***FIG***FIG***FIG***FIG***FIG***
\begin{figure*}
\includegraphics[width=1.05\textwidth]{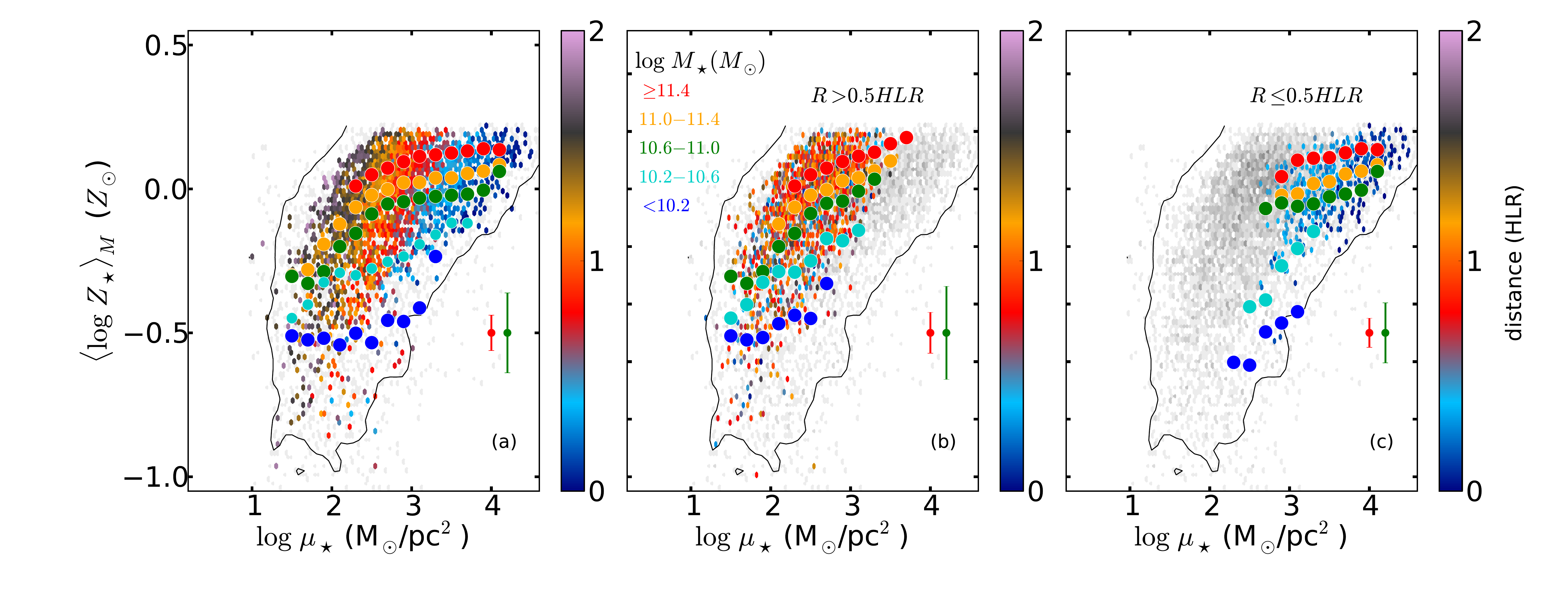}
\caption{{\em Left:} Local $\mu$ZR, color coded by $R$; circles show the ($\mu_\star$-binned) mean relations obtained by breaking up the sample into five $M_\star$ intervals. {\em Middle:} Same as (a), but restricting the analysis to spatial regions outwards of $R = 0.5$ HLR. {\em Right:} As the middle panel, but for the inner $R < 0.5$ HLR regions. Error bars represent the typical dispersion in metallicity in the local MZR in two of the five $M_\star$ intervals.
  }
\label{fig:muZR2}
\end{figure*}
%***FIG***FIG***FIG***FIG***FIG***FIG***FIG***FIG***FIG***FIG***

The results reported above send mixed messages as to what the main driver of stellar metallicities might be: 
While Fig.\ \ref{fig:MZR1} indicates that $M_\star$ is involved in establishing global $Z_\star$ values, Fig.\ \ref{fig:muZR1} points to $\mu_\star$ playing a major role in defining the local $Z_\star$. It thus seems that {\em both local and global effects play relevant roles in determining the metallicity}.

Fig.\ \ref{fig:muZR2} shows that this is indeed the case. The dots in all panels repeat the local $\mu$ZR, but now breaking the mean (i.e., $\mu_\star$-binned) relation (large circles) into five $\sim$ equally populated $M_\star$ intervals. Panel (a) shows results considering all 6000 radial bins, while panels (b) and (c) take into account only points outside and inside $R = 0.5$ HLR, respectively. The figure thus explores the whole $M_\star$-$\mu_\star(R)$-$Z_\star(R)$-$R$ landscape.

Comparing Fig.\ \ref{fig:muZR2}a with the mean $\mu$ZR traced by gray circles in Fig.\ \ref{fig:muZR1} one immediately realizes that a pure $\mu$ZR  analysis overlooks the important role of $M_\star$ in defining the metallicity scale. Mass is a major source of scatter in the $\mu$ZR, with massive galaxies being locally (and globally) more chemically enriched than low mass ones for the same $\mu_\star(R)$. Still, the rise of $Z_\star(R)$ with $\mu_\star(R)$ for constant $M_\star$ shows the relevance of local effects, particularly for low and intermediate $M_\star$ systems.

Though revealing, the $M_\star$-dependent mean $\mu$ZRs in Fig.\ \ref{fig:muZR2}a do not distinguish the whereabouts within a galaxy, as disk and bulge/spheroid locations are treated on equal footing. A coarse way to discriminate these morphological components is to define them by a $R >$ or $< 0.5$ HLR criterion. Panel $b$ in Fig.\ \ref{fig:muZR2} shows the $\mu$ZR$(M_\star)$ for $R > 0.5$ HLR. Except for the 41 ellipticals, this plot should be read as a $\mu$ZR$(M_\star)$ for galactic disks. In stark contrast to Fig.\ \ref{fig:muZR2}a, no downwards flattening is seen. Stellar metallicities increase steadily with $\mu_\star$, modulated by a $M_\star$-related amplitude.

The panorama changes in Fig.\ \ref{fig:muZR2}c, where only the inner 
 zones are considered. For the three largest $M_\star$ ranges ($> 10^{10.6} M_\odot$), one sees 
a very weak dependence of $Z_\star$ on $\mu_\star$. Local effects thus seem of secondary importance in the dense regime of the inner regions of massive galaxies. The $M_\star < 10^{10.6} M_\odot$ bins, however, still follow a strong ($M_\star$-modulated) $\mu$ZR. Galaxies in these low $M_\star$ bins are predominantly (87\%) late type spirals (Sc--Sd), with small or non existent bulges (possibly resulting from secular disk evolution; Fisher \& Drory 2011), which explains their disk-like $\mu$ZR.

\section{Discussion}

The general picture one draws from these results is that 
both global ($M_\star$-driven) and local ($\mu_\star$-driven) effects are important in determining the stellar (and, by extension, also the nebular) metallicities in galaxies, and that the overall balance between these two varies with the location within a galaxy. Two clearly distinct regimes are identified: {\em (i)} One related to disks, where $\mu_\star$  (hence local physics) regulates metallicity, modulated by an $M_\star$-related amplitude, and {\em (ii)} the other pertaining to the bulge/spheroid component, where $M_\star$ 
dominates the physics of star formation and chemical enrichment.

These results tie in nicely with the recent analysis of GD14. We have shown there that mean stellar ages (a first moment descriptor of the SFH) relate strongly to $\mu_\star$ in galactic disks, indicating that local properties dictate the pace of star-formation. The slower growth (hence younger ages) found at low $\mu_\star$ should lead to less metal enrichment, in agreement with our $\mu$ZR. Within bulges/spheroids, $M_\star$ is a much more relevant driver of the SFH. Most of the star formation in these regions was over long ago, leading to fast metal enrichment and little or no chemical evolution since those early days, as found in this Letter.

Despite the overall consistency and physical appeal of our results, there are some caveats. Aside from limitations of the method (e.g., Cid Fernandes et al.\ 2014), and the choice of $R=0.5$ to segregate bulges from disks, a more philosophical kind of caveat holds the promise of interesting future work. The goal of this Letter was to evaluate the role of global and local effects in determining stellar metallicities in and within galaxies. To this end, we have chosen $M_\star$ and $\mu_\star$ as the representatives of global and local effects, respectively, but was this a wise choice of variables? 

Our answer at this stage is a``yes'' for $M_\star$, but only a ``perhaps'' for $\mu_\star$. Both the SFH of a galaxy and its chemical evolution are thought to be affected by its mass, including its dark $+$ gaseous $+$ stellar components. 
In practice, $M_\star$ is the most readily available proxy for total mass, and thus a relevant variable. 

The status of $\mu_\star$ as a tracer of local effects is less clear. The choice of $\mu_\star$ as an independent variable is inspired by Schmidt (1959), who postulates a density controlled SFR law, and by the fact that our analysis provides robust estimates of $\mu_\star$. Clearly, $\mu_\star$ is but a 
proxy for complex star formation physics. The fact that we find $Z_\star$ to correlate strongly with $\mu_\star$ in disks suggests it to be a reasonable empirical proxy for local effects. Yet, the fact that the disk $\mu$ZR has its amplitude modulated by $M_\star$ raises doubts as to how fundamental $\mu_\star$ really is.

While one can think of ways to relate local $Z_\star$ values to the global $M_\star$, there are other ways of formulating the problem. The $Z_\star$-scaling role of $M_\star$ in the disk $\mu$ZR may be reflecting the works of a different $M_\star$-related property, like morphological type, gas fraction, in/out-flow rates, or others. Future work should explore these possibilities.

\begin{acknowledgements} 
This Letter is based on data obtained by the CALIFA survey (http://califa.caha.es), funded by the Spanish MINECO
grants ICTS-2009-10, AYA2010-15081, and the CAHA operated jointly by the Max-Planck IfA and the IAA (CSIC). The CALIFA Collaboration thanks the CAHA staff for the dedication to this project. 
Support from CNPq (Brazil) through Programa Ci\^encia sem Fronteiras (401452/2012-3) is duly acknowledged. AG acknowledges financial support from the European Union Seventh Framework Programme (FP7/2007-2013) under grant agreement n. 267251 (AstroFIt). CJW acknowledges support through the Marie Curie Career Integration Grant 303912.

\end{acknowledgements}

\end{document}